\documentclass[twoside,fleqn]{article}
\usepackage{espcrc2}

\usepackage{feynmp}
\usepackage{amsmath}
\usepackage{mcite}
\newcommand{\ii}{\mathrm{i}}
\newcommand{\braket}[1]{\mathinner{\langle{#1}\rangle}}
\newcommand{\sbraket}[1]{\lbrack #1\rbrack}

\newcommand{\ket}[1]{\mathinner{|{#1}\rangle}}
\makeatletter
\newcommand{\fmslash}[2][0mu]{%
  \mathchoice
    {\fmsl@sh\displaystyle{#1}{#2}}%
    {\fmsl@sh\textstyle{#1}{#2}}%
    {\fmsl@sh\scriptstyle{#1}{#2}}%
    {\fmsl@sh\scriptscriptstyle{#1}{#2}}}
\newcommand{\fmsl@sh}[3]{%
  \m@th\ooalign{$\hfil#1\mkern#2/\hfil$\crcr$#1#3$}}
\makeatother

\newcommand{\nofrac}[2]{\genfrac{}{}{0pt}{1}{#1}{#2}}
\newcommand{\sm}{\scriptstyle}
\begin{fmffile}{mhv_figs}
\fmfset{arrow_len}{2.5mm}
\allowdisplaybreaks

\title{Born amplitudes in QCD from scalar diagrams\thanks{Talk given at QCD 05:
 12th International QCD Conference, 4-9 Jul 2005, Montpellier, France }}

\author{Christian Schwinn\address{Institut f\"ur Theoretische Physik E,
RWTH Aachen, D-52056 Aachen, Germany}
 and Stefan Weinzierl\address{Institut f\"ur Physik, 
   Johannes-Gutenberg-Universit\"at, Staudingerweg 7,  D-55099 Mainz, Germany}%
         }

\begin{document}

\begin{abstract}
We review recent developments for the calculation of Born amplitudes
in QCD.  This includes the computation of gluon helicity amplitudes
from MHV vertices and an approach based on scalar propagators and a
set of three- and four-valent vertices.  The latter easily generalizes
to amplitudes with any number of quark pairs.  The quarks may be
massless or massive.
\end{abstract}

\maketitle

\section{Introduction}
\label{sect:intro}

At the CERN Large Hadron Collider (LHC) events with six or more
partons in the final state will occur with significant rates and
contribute to the background for electroweak physics and for searches
for physics beyond the standard model.  Already at tree level, the
number of  Feynman diagrams contributing to the relevant scattering amplitudes
grows rapidly with the number
of external legs,
for instance over $34 000$ diagrams contribute to
the eight gluon amplitude at Born level.  
More efficient methods to calculate multiparton 
scattering amplitudes have been developed, employing
 a decomposition of QCD Born amplitudes into
gauge-independent color structures, the spinor helicity formalism and
recursive techniques \cite{Mangano:1990by,*Dixon:1996wi}.  Using such
methods it has been shown that helicity amplitudes for specific
helicity combinations have a remarkably simple analytic form or
vanish altogether \cite{Parke:1986gb,Berends:1987me}.  In particular,
a compact analytical form is known for the first non-vanishing
amplitude---the so called maximal-helicity violating amplitude (MHV
amplitude)---where two gluons have the opposite helicity of the
remaining ones.

Recently, new methods for further simplifications in the calculation
of multiparticle amplitudes have been developed, initiated by
insight gained from the transformation of the amplitudes to twistor
space and a conjectured relation to a certain string model
\cite{Witten:2003nn}.  This development led to a construction by
Cachazo, Svr\v{c}ek and Witten (CSW) \cite{Cachazo:2004kj} that
expresses helicity amplitudes with arbitrary helicity configurations
in terms of diagrams with vertices given by off-shell continuations of
MHV-amplitudes, connected by scalar propagators. For certain helicity
configurations,  this formalism has led to much simpler expressions  
than the conventional expansion in terms of Feynman diagrams.

In this talk we will briefly review these recent developments
and introduce a method~\cite{SW:MHV} to compute
Born amplitudes in QCD---including also massless
or massive quarks---from scalar propagators and a set
of three- and four-valent vertices.

In section~\ref{sec:helicity} we recall the use of helicity methods and the
color decomposition in the computation of born amplitudes. In
section~\ref{sec:csw} we review the CSW construction for QCD amplitudes in
terms of MHV vertices.  In section~\ref{sec:scalar} we introduce the formalism
of scalar Feynman rules in QCD before discussing the extension to massive
quarks in section~\ref{sec:massive}.
\section{Helicity methods in QCD}
\label{sec:helicity}
An efficient algorithm for  calculations in QCD 
starts with the  separation of the color structure from the
 the kinematic and spinor
structure of born amplitudes. This is accomplished using the so called
\emph{color decomposition}.
In the pure gluonic case tree level amplitudes with $n$ external gluons may be written in the form
\begin{multline}
{\cal A}_{n}(1,2,...,n) \\
=  g^{n-2} \sum_{\sigma  \in S_{n}/Z_{n}} 2 \; \mbox{Tr} \left(
 T^{a_{\sigma(1)}} ... T^{a_{\sigma(n)}} \right)\\
 A_{n}\left( \sigma(1), ..., \sigma(n) \right), 
\end{multline}
where the sum is over all non-cyclic permutations of the external gluon legs.
The color structure is contained in group theoretical factors given by
 traces over the generators
$T^a$ of the fundamental representation of $SU(3)$.
The quantities $A_n(\sigma(1),...,\sigma(n))$, called the \emph{partial
  amplitudes}, contain the kinematic information and are separately gauge
invariant.
They are color-ordered, e.g. only diagrams with a particular cyclic ordering
of the gluons contribute. For amplitudes containing external quarks,
similar decompositions exist.

The calculation of the 
partial amplitudes is simplified by the spinor helicity formalism~\cite{Berends:1981rb,*DeCausmaecker:1981bg,*Gunion:1985vc,*Xu:1986xb}.
The basic ingredients are two-component Weyl spinors
$\ket{p +}$ and $\ket{p -}$ that are defined as solutions of the Weyl equations
\begin{equation}
  \sigma^\mu p_\mu \ket{p -}=0\quad,\quad  \bar\sigma^\mu p_\mu \ket{p +}=0
\end{equation}
where $\sigma^\mu=(1,-\vec \sigma)$ and  $\bar \sigma^\mu=(1,\vec \sigma)$.
Antisymmetric spinor products can be defined by
\begin{equation}
  \braket{pk}=\braket{p-|k+}\quad,\quad  \sbraket{pk}=\braket{p+|k-}
\end{equation}
Gluons can be included in the spinor formalism by expressing the 
polarization vectors of the external gluons in terms
of Weyl-Spinors:
\begin{equation}
\epsilon^\pm_\mu(k,q)=\pm \frac{\braket{q\mp|\gamma_\mu|k\mp}}{\sqrt 2 \braket{q\mp |k\pm}} 
\end{equation}
where $q$ is an arbitrary light-like reference momentum. On can show that
different choices of $q$ correspond to different gauge choices for the
external gluons.

The helicity formalism allowed to derive remarkably
simple analytic formula for  helicity amplitudes with
 specific helicity combinations~\cite{Parke:1986gb,Berends:1987me}.
In  particular, the pure gluon amplitude vanishes if all gluons have the same helicity, or if all gluons except one
have the same helicity.
The first non-vanishing gluon amplitude is 
obtained if $n-2$ gluon have one type of helicity, and $2$ gluons the other
type. This so called MHV amplitude is given by
the \emph{Parke-Taylor formula}~\cite{Parke:1986gb}
\begin{multline}
\label{eq:mhv}
  A_n(1^+,\dots,i^-,\dots, j^-,\dots n^+)\\
=\ii 2^{n/2-1}
\frac{\braket{ij}^4}{\braket{12}\braket{23}\dots\braket{(n-1)n}\braket{n1}}
\end{multline}
This was proven later by Berends and Giele~\cite{Berends:1987me} using 
recursive techniques.
For reviews of further developments and applications of the helicity formalism 
see e.g.~\cite{Mangano:1990by,*Dixon:1996wi}.

\section{QCD amplitudes from MHV diagrams}
\label{sec:csw}
The fact, that the MHV amplitudes for an arbitrary number of
external gluons take the simple form~\eqref{eq:mhv} motivated a number
of authors to search for an explanation outside the usual Feynman diagrammatic
methods~\cite{Nair:1988bq,*Bardeen:1995gk,*Cangemi:1996rx,*Chalmers:1996rq}. Recently Witten related scattering amplitudes of
(super-) Yang-Mills theory to a certain 
string model in Twistor space~\cite{Witten:2003nn}. 
Motivated by this connection, 
on the field theoretic
side Cachazo, Svr\v{c}ek and Witten~\cite{Cachazo:2004kj} subsequently
proposed a construction of  \emph{all} tree-level QCD amplitudes in terms
of MHV vertices~\eqref{eq:mhv} as building blocks. This construction allows
a considerable
 reduction of the computational effort for multiparton amplitudes
compared to the usual expansion in terms of Feynman diagrams.

As a first step, an 
appropriate off-shell continuation of the MHV amplitudes has to be
introduced. Since a spacelike or timelike four momentum cannot be decomposed
into just two Weyl spinors, the formula~\eqref{eq:mhv} cannot be used
for off-shell momenta in a
straightforward way. 
Here we chose the prescription of~\cite{Kosower:2004yz}
that is slightly different from the original CSW prescription. 
To translate an off-shell momentum into spinor language, one 
decomposes an arbitrary off-shell momentum into two
lightlike vectors according to
\begin{equation}
\label{eq:momentum}
  k^\mu = k^{\flat\mu}+\frac{k^2}{2(k\cdot q)} q^\mu
\end{equation}
where $q^\mu$ is a fixed, lightlike reference four-vector.
We can now associate two-component spinors $\ket{k^\flat+}$ and
$\ket{k^\flat -}$ with the \emph{projected momentum}
$k^{\flat\mu}$.
An MHV vertex including an off-shell gluon with momentum
 $k_l$ can now be defined 
using the spinor corresponding to the projected momentum $k_l^\flat$,
i.e. by replacing $\braket{ln}\to\braket{l^\flat n}$ 
in the expression~\eqref{eq:mhv}. If all external gluons are off-shell,
the MHV vertex is hence given by
\begin{multline}
  V_{\text{CSW}}(1^+,\dots,i^-,\dots, j^-\dots,n^+)\\
 =A(1^{\flat +},\dots,i^{\flat -},
  \dots, j^{\flat -}\dots,n^{\flat +})
\end{multline}
The CSW prescription then expresses an arbitrary QCD amplitude with
$n^-$ external negative helicity gluons in terms of $d=n^- -1$ MHV
vertices, where scalar propagators connect $+$ and $-$ labels. 

As an example, consider the so called next-to maximal helicity violating
 (NMHV) amplitudes with three gluons of negative helicity. For simplicity, 
we consider the amplitudes
 $A(1^-,2^-,3^+,\dots n^-)$ where the three negative helicity gluons
are adjacent. The algorithm begins by
 distributing the three external negative helicity gluons in all
possible ways over two MHV vertices. For our example,
there are three partitions $\left((1,2),n\right)$,
 $\left((1,(2,n)\right)$ and $\left(2,(n,1)\right)$. 
Next, connect the vertices
by a scalar propagator and distribute the positive helicity gluons in all possible
ways compatible with the cyclic color ordering. 
In the present example, 
the combination $\left((1,(2,n)\right)$ gives a vanishing contribution so
 only two topologies contribute.
The amplitude is given by a sum over the possible insertions of the
positive helicity gluons:
 \begin{multline}
A(1^-,2^-,3^+,\dots n^-)=\\
\sum _{j=3}^{n-1} \qquad
  \parbox{30mm}{
  \fmfframe(2,2)(2,2){
  \begin{fmfgraph*}(23,18)
    \fmfleftn{l}{4}
    \fmfrightn{r}{4}
       \fmf{plain}{l1,v1,l2}
    \fmf{plain}{r1,v2,r2}
    \fmf{plain,label=$\sm +-$,tension=2}{v1,v2}
    \fmf{plain}{l4,v1,l3}
    \fmf{plain}{r4,v2,r3}
    \fmfv{d.sh=circle,d.fi=empty,d.si=5mm}{v1,v2}
     \fmfv{label=$\sm{1^-}$,la.di=2pt}{l1}
     \fmfv{label=$\sm{2^-}$,la.di=2pt}{l2}
     \fmfv{label=$\sm{j^+}$,la.di=2pt}{l4}
     \fmfv{label=$\sm{(j+1)^+}$,la.di=2pt}{r4}
     \fmfv{label=$\sm{n^-}$,la.di=2pt}{r1} 
    \end{fmfgraph*}}}  \\[0.2cm]
+\quad
  \parbox{30mm}{
  \fmfframe(2,2)(2,2){
  \begin{fmfgraph*}(16,23)
    \fmfbottomn{l}{4}
    \fmftopn{r}{4}
       \fmf{plain}{l1,v1,l2}
    \fmf{plain}{r1,v2,r2}
    \fmf{plain,label=$\nofrac{-}{+}$,tension=2}{v1,v2}
    \fmf{plain}{l4,v1,l3}
    \fmf{plain}{r4,v2,r3}
    \fmfv{d.sh=circle,d.fi=empty,d.si=5mm}{v1,v2}
     \fmfv{label=$\sm{1^-}$,la.di=2pt}{l1}
     \fmfv{label=$\sm{n^-}$,la.di=2pt}{l2}
     \fmfv{label=$\sm{(j+1)^+}$,la.di=2pt}{l4}
     \fmfv{label=$\sm{j^+}$,la.di=2pt}{r4}
     \fmfv{label=$\sm{2^-}$,la.di=2pt}{r1} 
  \end{fmfgraph*}}}
\end{multline}
In this way, the $n$ gluon NMHV amplitude is expressed in terms of $2(n-3)$
CSW diagrams, reducing the growth of computational effort with the number
of external particles drastically compared to the usual expansion in terms of 
Feynman diagrams.
A closed expression for all NMHV amplitudes for gluons---including an
arbitrary placement of the three negative helicity gluons---has been
found using the CSW formalism in~\cite{Kosower:2004yz}.

Based
on so called on-shell recursion relations~\cite{Britto:2004ap} both
indirect arguments~\cite{Britto:2005fq} for the validity of the CSW rules
and a direct
derivation~\cite{Risager:2005vk} have been  given. The relation to  ordinary
Feynman diagrams, however, remains to be clarified.  
For further applications of the CSW formalism and extensions to fermions see 
e.g.~\cite{Georgiou:2004wu,*Georgiou:2004by,*Wu:2004fb,*Wu:2004jx,*Luo:2004ss,*Bena:2004ry}.
The application to loop diagrams has been initiated in~\cite{Cachazo:2004zb,*Cachazo:2004by,*Brandhuber:2004yw,*Luo:2004ss,*Bena:2004xu}.
For a review and references
to subsequent developments and the connection to twistor string theory
see e.g.~\cite{Cachazo:2005ga}.

\section{Scalar Feynman rules for QCD}
\label{sec:scalar}
So far, the CSW construction described in the previous section is applicable
only for massless quarks. Also, an understanding in terms of conventional
Feynman diagrams would be desirable. As a first step towards these issues, 
in~\cite{SW:MHV} we have introduced
 scalar diagrammatic rules for QCD partial amplitudes, that use a similar
language as the CSW construction.
As all propagators are scalars no contraction of Lorentz- or spinor-indices is present any more.
This makes our method also  well suited for a 
fast implementation on a computer.

We employ the decomposition of 
an off-shell momentum into two
lightlike vectors according to~\eqref{eq:momentum}.
In the following, we use spinor products defined by:
\begin{equation}
  \braket{ab}=\braket{a^\flat-|b^\flat+}\,,\, 
  \sbraket{ab}=\braket{a^\flat+|b^\flat-}
\end{equation}
We define the off-shell continuation of polarization vectors for
gluons using the Weyl-spinors $\ket{k^\flat}$ associated with the
projection $k^{\flat\mu}$ of the off-shell gluon momentum:
\begin{equation}
\label{eq:os-pol}
  \epsilon^\pm_\mu(q)=\pm\frac{\braket{q\mp|\gamma_\mu|k^\flat\mp}}
  {\sqrt 2 \braket{q\mp| k^\flat\pm}} 
\end{equation}
These definitions allow to rewrite the gluon propagator in the light-cone
gauge as a sum over the polarization vector~\eqref{eq:os-pol} 
and an additional instantaneous term:
\begin{multline}
\label{eq:g-prop}
\frac{\ii}{k^2}
\left(-g_{\mu\nu}+\frac{k_\mu q_\nu+k_\nu q_\mu}{q\cdot k}\right)\\
=
\frac{\ii}{k^2}\sum_{\sigma=\pm}\epsilon_\mu^\sigma(k^\flat,q)\epsilon_\mu^{*\sigma}(k^\flat,q)+\frac{\ii}{(q\cdot k)^2}q_\mu q_\nu  
\end{multline}
As noted earlier~\cite{Kosower:1989xy,Kosower:2004yz}, the 
instantaneous term can be absorbed into a redefinition of the
four-gluon vertex, without generating vertices with more than four
gluons. 
Similar results can also be obtained by eliminating the unphysical degrees
of freedom in the Lagrangian using the equations of motion~\cite{Chalmers:1998jb,*Siegel:1999ew}.

Contracting the polarization vectors with the vertices, one
obtains diagrammatic rules involving only scalar propagators $\ii/k^2$
connecting opposite helicities and vertices obtained by contracting
the standard Feynman rules with the polarization vectors. We call
these vertices "primitive vertices" to distinguish them from the
vertices of the standard Feynman rules on the one hand, and from the
MHV-vertices on the other hand.
The nonvanishing three gluon vertices obtained in this way are given 
by~\cite{SW:MHV}
\begin{equation}
\label{eq:3g}
  \begin{aligned}
    V_3(1^-,2^-,3^+) & = & 
\ii \sqrt{2} \frac{\braket{1 2}^4}{\braket{12}\braket{23} \braket{31}}
 \\
 V_3(1^+,2^+,3^-) & = & 
 \ii \sqrt{2} \frac{\sbraket{21}^4}{\sbraket{32} \sbraket{21}\sbraket{13}}
  \end{aligned}
\end{equation}
and cyclic permutations thereof. The modified four gluon vertices can be
found in~\cite{SW:MHV}.

To gain insight in the structure of scalar diagrams obtained using
the rules defined in this section, one can define 
the degree of a vertex or of an amplitude 
as the number of ``-''-labels minus one.
In the diagrammatic rules, only primitive vertices of degree zero and one 
occur.
Furthermore, the degree of an amplitude is exactly the sum of the degrees of the primitive vertices~\cite{SW:MHV}.
Therefore, one can see the structure of the
 CSW prescription emerge from our diagrammatic approach: 
a MHV amplitude is of degree one and contains exactly one primitive vertex of degree one, 
which is dressed up in all possible ways with degree zero vertices.
Similar, a pure gluon amplitude with three gluons of negative helicities is of degree two
and contains two vertices of degree
one, which again are combined in all possible ways with degree zero vertices.

\section{Scalar rules for massive quarks} 
\label{sec:massive}
Several methods exist to incorporate massive fermions into the helicity
formalism~\cite{Kleiss:1985yh,*Ballestrero:1994jn,*Dittmaier:1998nn,*vanderHeide:2000fx,Rodrigo:2005eu}. Here we will introduce an off-shell
continuation, using the same projection~\eqref{eq:momentum} as for the
gluon polarization vectors:
\begin{equation}\label{eq:os-spinors}
  \begin{aligned}
  u(+)&=\frac{\fmslash p+m}{\braket{p^\flat q}}\ket{q+}
 =\ket{p^\flat-}+\frac{m}{\braket{p^\flat q}}\ket{q+}\\
     u(-)&=\frac{\fmslash p+m}{\sbraket{p^\flat q}}\ket{q-}
  = \ket{p^\flat +} + \frac{m}{\sbraket{p^\flat q}} \ket{q-}
 \end{aligned}
\end{equation}
The normalization is chosen in order to allow for a smooth massless
limit.
The spinors~\eqref{eq:os-spinors} are eigenstates of
$\gamma^5\fmslash s$ with the spin vector~\cite{Kleiss:1985yh}
\begin{equation}
\label{eq:spin}
  s^\mu = \frac{p^\mu}{m}-\frac{m}{(p\cdot q)}q^\mu
\end{equation}
Therfore the reference momentum is not an unphysical quantity
that has to drop out in the final result for the helicity amplitude, as
in the case of the gluon polarization vectors, 
but rather defines the quantization axis of the quark spin.

Similarly to the case of the gluon propagator, the 
definitions~\eqref{eq:os-spinors}
allow to rewrite the quark propagator as a sum over off-shell spinors
and an additional instantaneous term:
\begin{equation}
\hspace*{-7mm}
  \frac{\ii(\fmslash p+m)}{p^2-m^2}=\frac{\ii}{p^2-m^2} \sum_\lambda u(-\lambda)\bar u(\lambda) + \ii \frac{\fmslash q}{2 (p\cdot q)}
\end{equation}
Again the instantaneous contribution can be absorbed by introducing
only quartic vertices~\cite{SW:MHV}, leading to additional
vertices involving two quarks and two gluons or four quarks.
In this way, also massive quarks can be described in terms of scalar
propagators $\ii/(k^2-m^2)$ connecting opposite helicities. 
The primitive vertices obtained by contracting the spinors into the
ordinary Feynman rules include vertices present both for massless
and massive quarks like
\begin{equation}
  \begin{aligned}
   V_3(1_q^+,2_{\bar{q}}^-,3^+)&= 
  \ii \sqrt 2 \frac{\sbraket{13}^2}{\sbraket{12}} \\
  V_3(1_q^-,2_{\bar{q}}^+,3^-)&= 
  \ii \sqrt 2 \frac{\braket{31}^2}{\braket{21}}
  \end{aligned}
\end{equation}
In addition, there are primitive
 vertices involving a helicity flip along the quark line
that vanish for massless quarks:
\begin{equation}
\label{eq:flip} 
 \begin{aligned}
    V_3(1_q^+,2_{\bar{q}}^+,3^-)&
    =\ii \sqrt 2 m \frac{\sbraket{12}^2}{\sbraket{23}\sbraket{31}}\\
     V_3(1_q^-,2_{\bar{q}}^-,3^+)& 
    =-\ii \sqrt 2 m \frac{\braket{12}^2}{\braket{23}\braket{31}}
  \end{aligned}
\end{equation}
The four-valent vertices can be found in~\cite{SW:MHV}.

In the amplitudes for massive quarks there are two new features
compared to the massless case.  First there are helicity
configurations, which vanish in the simultaneous massless and on-shell
limit, but remain non-zero in the on-shell limit for non-zero masses.
Examples of this kind are amplitudes with only positive helicity gluons:
\begin{equation}
\label{onshellforbidden}
 A_n\left(1_q^+, 2_{\bar{q}}^-, 3^+, ..., n^+ \right).
\end{equation}
Secondly, there are the helicity flip vertices~\eqref{eq:flip}
which are proportional to the mass and vanish therefore in the massless limit.
We note that $V_3(1_q^+,2_{\bar{q}}^+,3^-)$ is a degree zero vertex, whereas
$V_3(1_q^-,2_{\bar{q}}^-,3^+)$ has degree one.
This allows to determine the maximal number of flip-vertices in an amplitude
with a given degree.
Let us consider an amplitude with one massive quark pair. We first consider the helicity
configuration $q^+ \bar{q}^-$. Along the massive quark line we must have an equal number
of helicity flips induced by
$V_3(1_q^+,2_{\bar{q}}^+,3^-)$ and $V_3(1_q^-,2_{\bar{q}}^-,3^+)$.
Since the latter vertex is of degree one, the total number of helicity flips $f$
is bounded by the degree $d$ of the amplitude:
\begin{equation}
 f  \le  2 d  
\end{equation}
As an example it follows immediately 
from this bound that amplitudes of the form (\ref{onshellforbidden})
cannot contain a helicity flip.  
A similar argument applies to the helicity configuration $q^+ \bar{q}^+$.  
Here we have along the fermion line at least one helicity flip of degree zero 
together with at most $d$ additional pairs of degree one and zero helicity flips.
So the total number is bounded by 
\begin{equation}
 f  \le  2 d + 1.    
\end{equation}
For the helicity configuration $q^- \bar{q}^-$ there is at least one flip of
degree one, leaving at most $2(d-1)$ additional helicity flips so the
total number is bounded by 
\begin{equation}
 f \le  2 d - 1.   
\end{equation}

While a closed expression for amplitudes with a massive quark pair and
an arbitrary number of positive helicity gluons has been found
recently using a different choice of reference momenta for the massive
quarks~\cite{Rodrigo:2005eu}, we expect the formalism described in
this section and the classification of the amplitudes given above to
be useful for the calculation of amplitudes with massive quarks with
other helicity configurations.
 \section{Conclusions}
\label{sect:concl}
After a review of spinor helicity methods in QCD we have sketched
the CSW construction for expressing Born amplitudes in terms of 
MHV vertices, connected by scalar propagators.

We have then reviewed a method to calculate born amplitudes 
from a set of scalar, complex valued three- and four-valent vertices.
Our approach is not restricted to gluons only, but treats gluons, massless quarks and massive quarks
on almost equal footing.
It is possible to assign to each vertex a degree, given by the number of negative helicities minus one.
Only vertices of degree zero and one occur and the number of degree one
vertices in a diagram is equal to the number of MHV vertices in the CSW
construction.
In the case of massive quarks, it is possible to bound the number of 
helicity flips in terms of the degree of the amplitude.

We hope that our method will be useful to obtain closed expressions for
helicity amplitudes involving massive quarks and to get further insight
into the CSW construction. Work in this direction is in progress.
Our
method is also  well suited for a 
fast implementation on a computer.

\section*{Acknowledgments}
The work of CS has been supported
by the Deutsche Forschungsgemeinschaft through the
Gra\-du\-ier\-ten\-kolleg `Eichtheorien' at Mainz University.

\end{fmffile}
\bibliography{biblio}
\end{document}